\documentclass[twocolumn,showpacs,amsmath,amssymb,prl,superscriptaddress]{revtex4-1}
\usepackage{graphicx}
\usepackage{dcolumn}
\usepackage{bm}

\begin{document}

\title{Pairing-induced speedup of nuclear spontaneous fission}

\author{Jhilam Sadhukhan}
\affiliation{Department of Physics and Astronomy, University of Tennessee, Knoxville, Tennessee 37996, USA}
\affiliation{Physics Division, Oak Ridge National Laboratory, P. O. Box 2008, Oak Ridge, Tennessee 37831, USA}
\affiliation{Physics Group, Variable Energy Cyclotron Centre, 1/AF Bidhan Nagar, Kolkata 700064, India}
\author{J. Dobaczewski}
\affiliation{Institute of Theoretical Physics, Faculty of Physics, University of Warsaw, Pasteura 5, PL-02-093 Warsaw, Poland}
\affiliation{Department of Physics, P.O. Box 35 (YFL), University of Jyv\"askyl\"a, FI-40014  Jyv\"askyl\"a, Finland}
\affiliation{Joint Institute of Nuclear Physics and Applications, P. O. Box 2008, Oak Ridge, Tennessee 37831, USA}
\author{W. Nazarewicz}
\affiliation{Department of Physics and Astronomy and NSCL/FRIB Laboratory,
Michigan State University, East Lansing, Michigan  48824, USA}
\affiliation{Physics Division, Oak Ridge National Laboratory, P. O. Box 2008, Oak Ridge, Tennessee 37831, USA}
\affiliation{Institute of Theoretical Physics, Faculty of Physics, University of Warsaw, Pasteura 5, PL-02-093 Warsaw, Poland}
\author{J. A. Sheikh}
\affiliation{Department of Physics and Astronomy, University of Tennessee, Knoxville, Tennessee 37996, USA}
\affiliation{Physics Division, Oak Ridge National Laboratory, P. O. Box 2008, Oak Ridge, Tennessee 37831, USA}
\author{A. Baran}
\affiliation{Institute of Physics, University of M. Curie-Sk{\l}odowska, ul. Radziszewskiego 10, 20-031 Lublin, Poland}
\date{\today}

\begin{abstract}
\begin{description}
\item[Background]
Collective inertia is strongly influenced at the  level crossing at which
quantum system changes diabatically its microscopic configuration. Pairing
correlations tend to make the large-amplitude nuclear collective motion more adiabatic
by reducing the effect of those  configuration changes.
Competition between pairing and level crossing is thus expected to
have a profound impact on spontaneous fission lifetimes.

\item[Purpose]
To elucidate the role of nucleonic pairing on spontaneous fission, we
study the dynamic fission trajectories of $^{264}$Fm and  $^{240}$Pu
using the state-of-the-art self-consistent framework.

\item[Methods]
We employ the superfluid nuclear density functional theory with the
Skyrme energy density functional SkM$^*$ and a density-dependent
pairing interaction. Along with shape variables, proton and neutron
pairing correlations are taken as collective coordinates. The
collective inertia tensor is calculated within the nonperturbative
cranking approximation. The fission paths are obtained by using the
least action principle in a four-dimensional collective space of
shape and pairing coordinates.

\item[Results]
Pairing correlations are enhanced along the minimum-action fission
path. For the symmetric fission of $^{264}$Fm, where the effect of
triaxiality on the fission barrier is large, the geometry of fission
pathway in the space of shape degrees of freedom is weakly impacted
by pairing. This is not the case for $^{240}$Pu where pairing
fluctuations restore the axial symmetry of the dynamic fission
trajectory.

\item[Conclusions]
The minimum-action fission path is strongly impacted by nucleonic
pairing. In some cases, the dynamical coupling between shape and
pairing degrees of freedom can lead to a dramatic departure from the
static picture. Consequently, in the dynamical description of
nuclear fission, particle-particle correlations should be considered
on the same footing as those associated with shape degrees of
freedom.

\end{description}

\end{abstract}



\maketitle

{\it Introduction} --- Nuclear fission is a fundamental phenomenon
that is a splendid example of  a large-amplitude  collective motion
of a  system in  presence of many-body  tunneling. The corresponding
equations involve potential, dissipative,  and inertial terms
\cite{(Swi72)}. The individual-particle motion gives rise to shell
effects that influence the fission barriers and shapes on the way to
fission, and also strongly impact the inertia tensor through the crossings of
single-particle levels and resulting configuration changes
\cite{(Hil53),(Wil59),(Has78)}. The residual interaction between
crossing configurations is strongly affected by nucleonic pairing:
 the larger pairing gap $\Delta$ the more adiabatic is the collective
motion \cite{(Bra72),(Sch75),*(Sch78),(Str77),(Naz93c),(Nak98)}.

The enhancement of pairing correlations along the fission path was
postulated in early Ref.~\cite{(Mor74)} using simple physical
arguments. Since the collective inertia roughly depends on pairing
gap as $\Delta^{-2}$ \cite{(Uri66),(Bra72),(Led73),(Laz87),(Pom07)},
by choosing a pathway with larger $\Delta$, the fissioning nucleus
can lower the collective action. This means that in searching for the
least action trajectory the gap parameter should be treated as a
dynamical variable. Indeed, macroscopic-microscopic studies
\cite{(Sta85),(Sta89),(Loj99),(Mir10)} demonstrated that pairing
fluctuations  can significantly reduce the collective action; hence,
affect  predicted spontaneous fission lifetimes.

Our long-term goal is  to describe spontaneous fission (SF) within
the superfluid nuclear density functional theory  by minimizing the
collective action  in many-dimensional collective space. The
important milestone  was a recent paper \cite{(Sad13)}, which
demonstrated that  predicted SF pathways strongly depend on the
choice of the collective inertia and  approximations involved in
treating level crossings. The main objective of the present work is to
elucidate the role of nucleonic pairing on  SF by studying the
dynamic fission trajectories of $^{264}$Fm  and  $^{240}$Pu in a
four-dimensional collective space. In addition to  two quadrupole
moments defining the elongation and triaxiality of nuclear shape we
consider the strengths of  neutron and proton pairing fluctuations.
Since in  our model the effect of triaxiality on the fission barrier
is larger for $^{264}$Fm  ($\sim$4\,MeV) \cite{(Sta11)} than for
$^{240}$Pu ($\sim$2\,MeV) \cite{(She09)}, by considering these two
cases we can study the interplay between pairing dynamics and
symmetry breaking  effects \cite{(Neg89),(Naz93c)}.

{\it Theoretical framework} ---
To calculate the SF half-life, we closely follow the formalism
described in Ref.~\cite{(Sad13)}. In the semi-classical
approximation, the SF half-life  can be written
as~\cite{(Bar81),(Bar78)} $T_{1/2}=\ln2/(nP)$, where $n$ is the
number of assaults on the fission barrier per unit time (here we
adopt the standard value of $n=10^{20.38}s^{-1}$) and
$P=1/(1+e^{2S})$ is the penetration probability expressed in terms of
the fission action integral,
\begin{equation}
\label{action-integral}
S(L)=\int_{s_{\rm in}}^{s_{\rm out}}\frac{1}{\hbar}\sqrt{2\mathcal{M}_{\text{eff}}(s)
\left(V(s)-E_0\right)}\,ds  ,
\end{equation}
calculated along the fission path $L(s)$. The effective inertia
$\mathcal{M}_{\text{eff}}(s)$  \cite{(Bar81),(Bar78),(Bar05),(Sad13)}
is obtained from the multi-dimensional nonperturbative cranking inertia tensor
$\mathcal{M}^C$. $E_0$ is the collective ground state energy, and
$ds$ is the element of length along $L(s)$. To compute the  potential
energy $V$, we subtract the vibrational zero-point energy
$E_{\text{ZPE}}$ from the Hartree-Fock-Bogoliubov (HFB) energy
$E_{\text{HFB}}$ obtained  self-consistently from  the  constrained
HFB equations for the Routhian:
\begin{equation}\label{energy}
\hat{H}'=\hat{H}_\text{HFB} -\sum_{\mu=0,2} \lambda_{\mu}\hat{Q}_{2\mu} -\sum_{\tau=n,p}\left(\lambda_\tau \hat{N}_\tau-\lambda_{2\tau}\Delta\hat{ N}^2_\tau\right),
\end{equation}
where $\hat{H}_\text{HFB}$, $\hat{Q}_{2\mu}$, and $\hat{N}_\tau$
represent the HFB hamiltonian, axial ($\mu=0$) and nonaxial ($\mu=2$)
components of the mass quadrupole moment operator, and neutron
($\tau=n$) and proton ($\tau=p$) particle-number operators, respectively.
The particle-number dispersion terms $\Delta\hat{ N}^2_\tau = \hat{
N}^2_\tau - \langle \hat{ N}_\tau\rangle^2$, controlled by the
Lagrange multipliers $\lambda_{2\tau}$, determine dynamic pairing
correlations of the system \cite{(Vaq11),(Vaq13)}. That is,
$\lambda_{2\tau}=0$ corresponds to the static HFB pairing. Dynamic
pairing fluctuations stronger than those obtained within the static solution are described by
$\lambda_{2\tau}>0$. The overall magnitude of pairing correlations
(static + dynamic) can be related to the average pairing gap
$\Delta_{\tau}$ \cite{(Dob84),(Dob95a)}.  In this study,
$\lambda_{2\tau}$ are used as two independent dynamical coordinates
to scan over a wide range of pairing correlations (or
$\Delta_{\tau}$).

The one-dimensional path $L(s)$ is defined in the multidimensional
collective space by specifying the  collective variables $\{X_i\}
\equiv \{Q_{20}, Q_{22}, \lambda_{2n}+\lambda_{2p},
\lambda_{2n}-\lambda_{2p}\}$ as functions of path's length $s$.
Furthermore, to render collective coordinates dimensionless, we
define $x_i=X_i/\delta q_i$. In this way,  $ds$ becomes
dimensionless as well. In this work, we take  $\delta q_1=\delta
q_2=1$\,b and $\delta q_3=\delta q_4=0.01$. The dynamical coordinates
$x_3$ and $x_4$  control respectively the isoscalar and isovector
pairing fluctuations.

As a continuation of our previous study ~\cite{(Sad13)}, we first
consider the SF of $^{264}$Fm, which is predicted to undergo a
symmetric split into  two doubly magic  $^{132}$Sn
fragments~\cite{Sta09}. Therefore, the crucial shape degrees of
freedom in this case are elongation and triaxiality; they are
represented by quadrupole moments $Q_{20}$ and $Q_{22}$ defined as in
Table 5 of Ref.~\cite{(Dob04b)}. To compute the total energy
$E_{\text{HFB}}$ and inertia tensor $\mathcal{M}^C$, we employed the
symmetry-unrestricted  HFB solver HFODD
(v2.49t)~\cite{(Sch12)}. To be consistent with  the previous
work ~\cite{(Sad13)}, we use the Skyrme energy density functional
SkM$^*$ \cite{(Bar82)} in the particle-hole channel.

The particle-particle interaction  is approximated by the
density-dependent mixed pairing force \cite{doba02}. The zero-point
energy $E_\text{ZPE}$ is estimated by using the  Gaussian overlap
approximation~\cite{(Sta89),(Bar07),(Sta13)}. To obtain the expression
for $E_\text{ZPE}$, we neglected the derivatives of the pairing fields with
respect to $\lambda_{2\tau}$; we checked, however, that the topology
of fission path is hardly sensitive to the detailed structure of
$E_\text{ZPE}$.  The inertia tensor $\mathcal{M}^C$ was obtained from
the nonperturbative cranking approximation to Adiabatic Time
Dependent HFB as described in Refs.~\cite{(Bar11),(Sad13)}. The
density-matrix derivatives with respect to collective coordinates
used to compute $\mathcal{M}^C$~\cite{(Bar11)}, were obtained by using
finite differences with steps  $\delta q_i$. Finally, to obtain the
minimum action pathways we adopted two independent algorithms to
ensure the robustness of the result: the dynamical programing method
(DPM) ~\cite{(Bar81)} and the Ritz method~\cite{(Bar78)}. In all
cases considered, both approaches give consistent answers.

\begin{figure}[htb]
\includegraphics[width=\columnwidth]{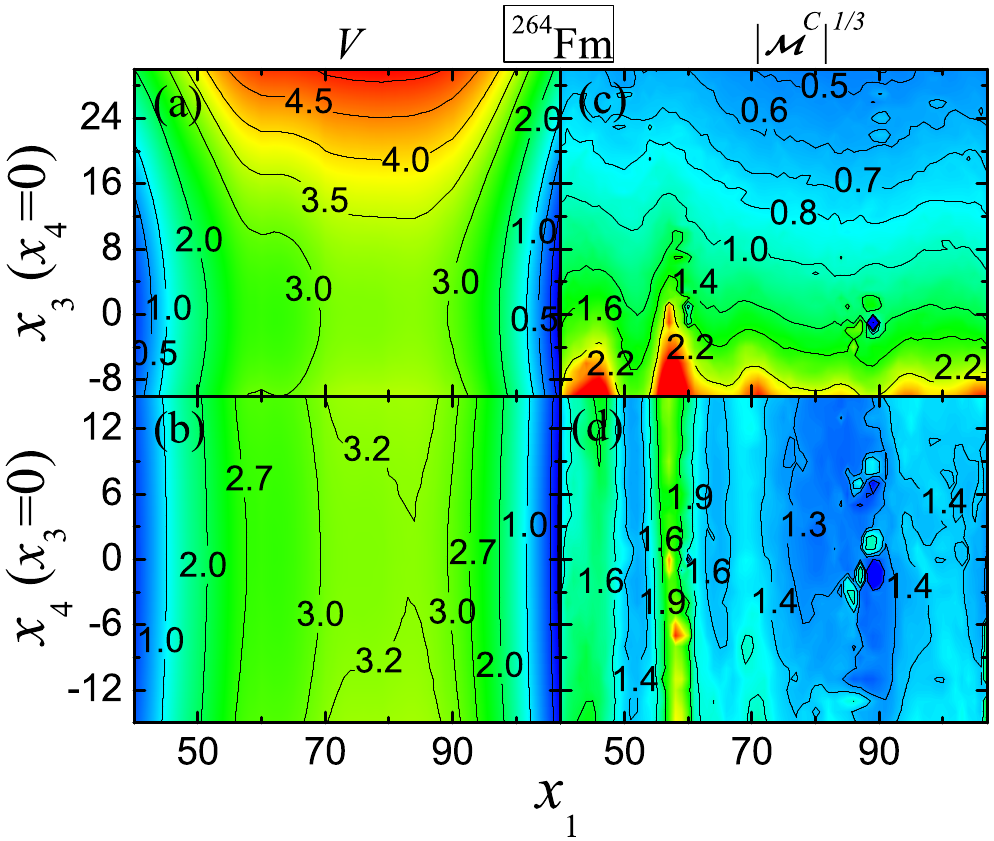}
\caption[C1]{\label{plot1}
(Color online) Contour maps of $V$ (left, in MeV) and
$|\mathcal{M}_C|^{1/3}$  (right, in $\hbar^2$\,MeV$^{-1}$/1000),
calculated for $^{264}$Fm in the $x_1$--$x_3$ plane for $x_4=0$ (top)
and $x_1$--$x_4$ plane for $x_3=0$ (bottom). The energies are plotted
relatively to the ground-state value.
}
\end{figure}

{\it Results} ---
In the first step, to assess the
relative importance of isoscalar and isovector pairing degrees of
freedom, the minimum-action path was calculated in the
three-dimensional space of coordinates $x_1$, $x_3$, and $x_4$. To this end, we adopted a 90$\times$61$\times$31 mesh with
$21\le x_1 \le 110$, $-10\le x_3 \le 50$, and $-15 \le x_4 \le 15$.
Coordinate $x_2$ was fixed according to the two-dimensional
dynamical path of  Ref.~\cite{(Sad13)}. The contour maps of $V$  in
the $x_1$--$x_3$ plane for $x_4=0$ and $x_1$--$x_4$ plane for $x_3=0$ are displayed in
Fig.~\ref{plot1} (left).

Since the individual components of the full (three-dimensional)
inertia tensor are difficult to interpret, following \cite{(Sad13)}
in Fig.~\ref{plot1} (right) we show the cubic-root-determinant of
inertia tensor $|\mathcal{M}_C|^{1/3}$. It can be seen that at large
values of $x_3$, the peaks in $|\mathcal{M}^C|^{1/3}$ due to level
crossings disappear and, moreover, the magnitude of inertia generally
decreases with $x_3$. This is consistent with general expectations
for the effect of paring on collective inertia. On the other hand,
variations in $x_4$ have little effect on $|\mathcal{M}^C|^{1/3}$ and
$V$. This result is confirmed by computing  the minimum action path
in the $(x_1,x_3,x_4)$ space: the fissioning system prefers to
maintain large proton and neutron pairing gaps and, at the same time,
$x_4 \approx 0$. Consequently, in the SF, this degree of freedom
seems to play less important role.

\begin{figure}[htb]
\includegraphics[width=1.0\columnwidth]{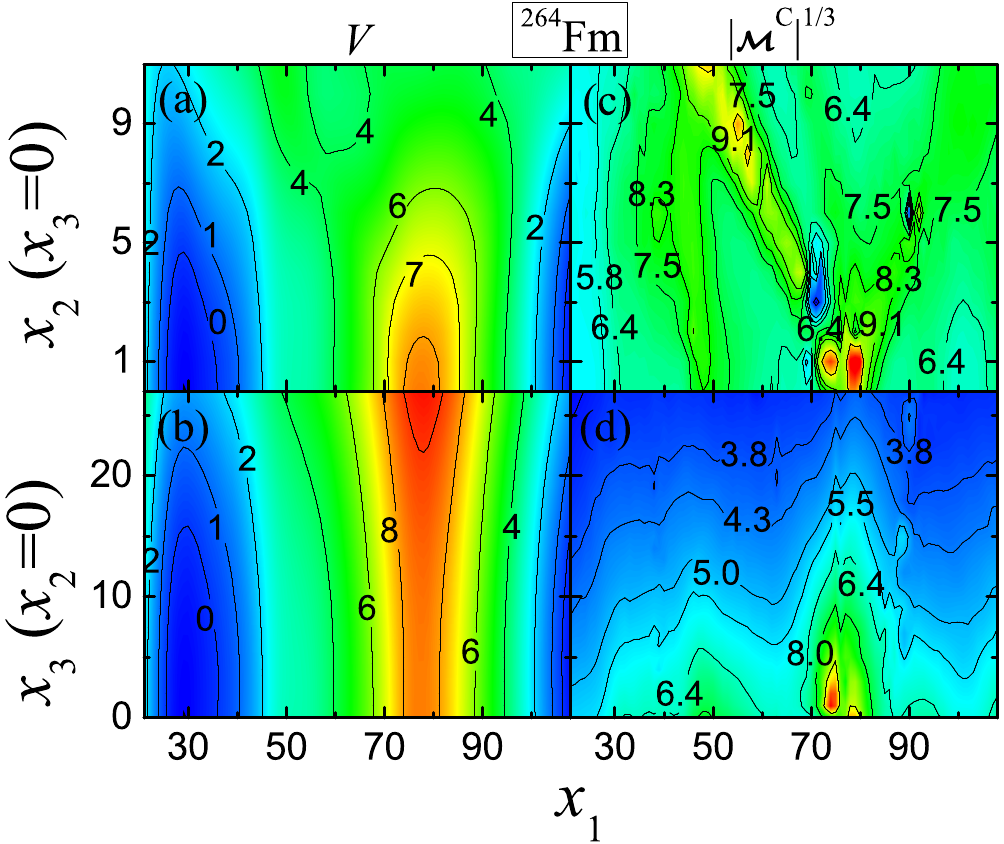}
\caption[C1]{\label{plot3}
(Color online) Similar as in Fig.~\ref{plot1} except for the
$x_1$--$x_2$ plane for $x_3=x_4=0$
(top) and
$x_1$--$x_3$ plane for $x_2=x_4=0$
(bottom).
}
\end{figure}
In the previous work, we have shown that the minimum action path
breaks the axial symmetry to avoid level crossings and minimize the
level density of single-particle states around the Fermi level. In
contrast,  pairing correlations grow with single-particle level
density. Consequently, pairing is expected to impact  $V$  and
$\mathcal{M}_{\text{eff}}$ in a different way. Namely, as pairing (or
$x_3$) increases, the potential energy is expected to grow -- as one
departs from the self-consistent value -- while the collective
inertia is reduced. The interplay between these two opposing
tendencies determines the least-action trajectory. To evaluate how
the fission path is modified due to pairing, in the next step we
minimize the collective action  in the $(x_1,x_2,x_3)$ space. Here we
assume $x_4=0$  and adopt the value of $E_0=1$\,MeV  to be consistent
with Ref.~\cite{(Sad13)}. Figure~\ref{plot3}  displays the resulting
contour maps of  $V$ and $|\mathcal{M}^C|^{1/3}$. The upper panels
correspond to the situation discussed in Ref.~\cite{(Sad13)}, in
which dynamical pairing is disregarded ($x_3=0$). As seen in
Fig.~\ref{plot3}(a),  triaxial coordinate $x_2$ reduces the fission
barrier height by slightly more than  4 MeV. The fluctuations seen in
$|\mathcal{M}^C|^{1/3}$ in Fig.~\ref{plot3}(c) reflect crossings of
single-particle levels at the Fermi level. The results shown in the
lower panels correspond to the axial shape ($x_2=0$); they  are
again consistent with the general dependence of potential energy and
collective inertia on pairing correlations.

\begin{figure}[htb]
\includegraphics[width=0.8\columnwidth]{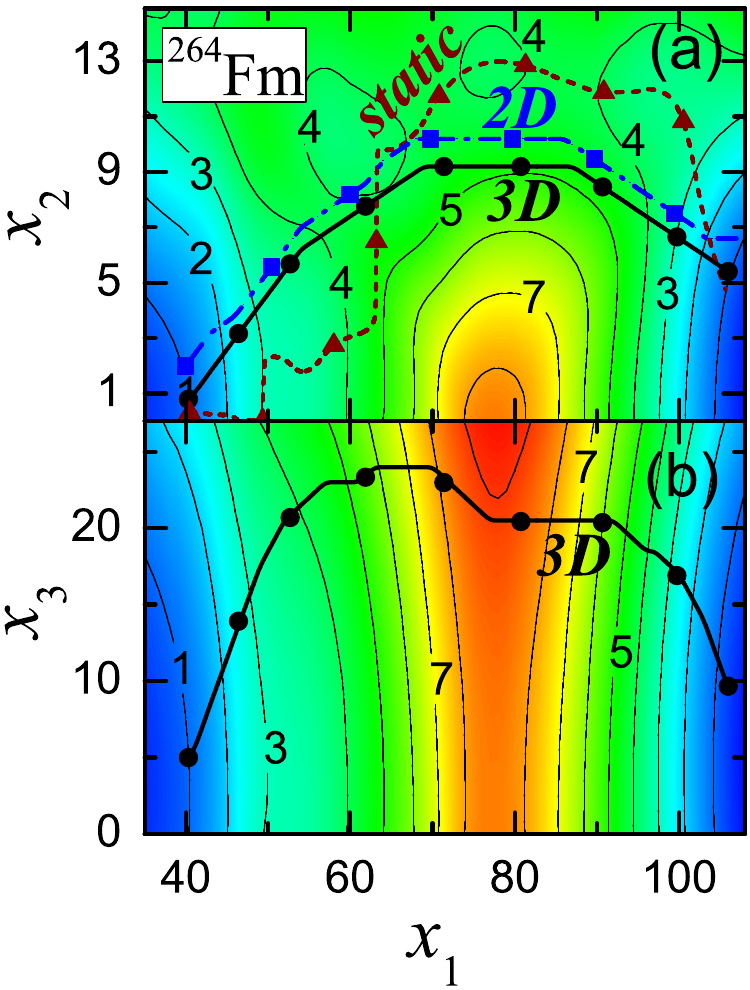}
\caption[C1]{\label{plot4}
(Color online) Projections of the three-dimensional (3D, solid line) dynamic SF
path for $^{264}$Fm on the $x_1$--$x_2$ plane for $x_3=x_4=0$ (a)
and $x_1$--$x_3$ plane for $x_2=x_4=0$ (b), calculated using the DMP
technique. The dash-dotted line shows for comparison the
two-dimensional (2D) path computed without pairing fluctuations. The static SF path corresponding to the minimized collective
potential \cite{(Sad13)} is also plotted (dotted line).
Symbols on the paths denote the path lengths in units of 10.
Potentials $V$ of Fig.~\protect\ref{plot3} are drawn as a background
reference.}
\end{figure}
In Figs.~\ref{plot4}(a) and \ref{plot4}(b), we show projections of
the minimum action path onto the $x_1$--$x_2$ and $x_1$--$x_3$
planes, respectively. The two-dimensional (2D) fission path
calculated without  pairing fluctuations ($x_3=x_4=0$) and
the static SF path corresponding to the valley of the minimized collective
potential are also shown
for comparison. Evidently, the triaxiality along the fission path 3D is
 reduced at the expense of enhanced pairing. Nevertheless,
owing to the reduced action $S$, the calculated SF
half-life of $^{264}$Fm in the 3D variant is decreased by as much as three decades.

\begin{figure}[htb]
\includegraphics[width=0.8\columnwidth]{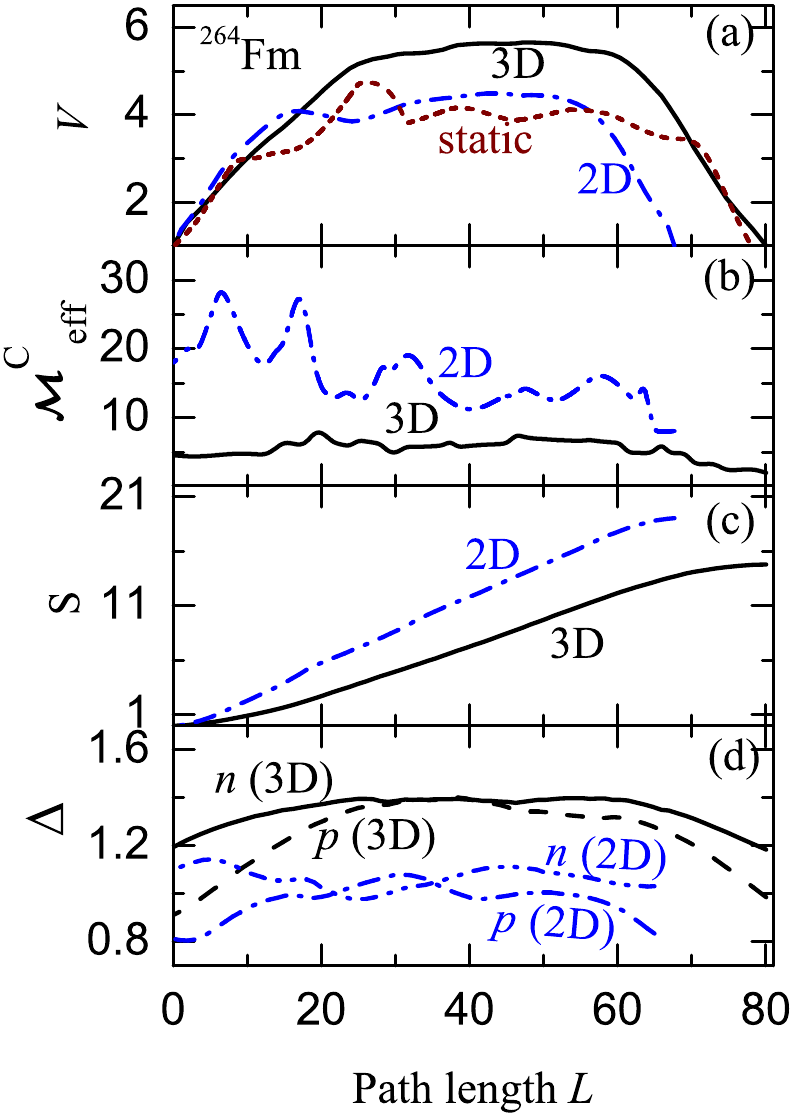}
\caption[C1]{\label{plot5}
(Color online) Potential $V$ (in MeV) (a),
effective inertia $\mathcal{M}^C_\text{eff}$ (in
$\hbar^2$\,MeV$^{-1}$/1000) (b), action $S$ (c), and average pairing
gaps $\Delta_n$ and $\Delta_p$ (in MeV) (d) plotted along the 2D (static pairing,
dotted line) and 3D (dynamic pairing, solid line) paths. The static fission barrier is  displayed for comparison in panel (a).
}
\end{figure}

Figure~\ref{plot5} summarizes our results for $^{264}$Fm. Namely, it
shows $V$,  $\mathcal{M}^C_\text{eff}$, $S$, and $\Delta_{\tau}$
along the fission paths calculated with dynamical (3D) and static (2D)
pairing. Compared to the 2D path, the
3D path is shorter and it favors lower collective inertia
at a cost of higher potential energy, both being the result of enhanced pairing
correlations.  It is interesting to notice that the collective potentials $V$
in 2D and 3D are fairly different, and they both deviate from the static result that is usually interpreted in terms of a fission barrier, or a saddle point.

While the least-action pathways in $^{264}$Fm are not that far from
the static SF path, this is not the case for $^{240}$Pu, where
the energy gain on the first barrier resulting from triaxiality is
around 2\,MeV, that is, significantly less than in $^{264}$Fm. To
illustrate the impact of pairing fluctuations on the SF of $^{240}$Pu, we
consider the least-action collective path between its ground state
and superdeformed fission isomer. In this region of collective space,
reflection-asymmetric degrees of freedom are less important; hence,
the 3D space of ($x_1, x_2, x_3$) is adequate.

\begin{figure}[htb]
\includegraphics[width=0.8\columnwidth]{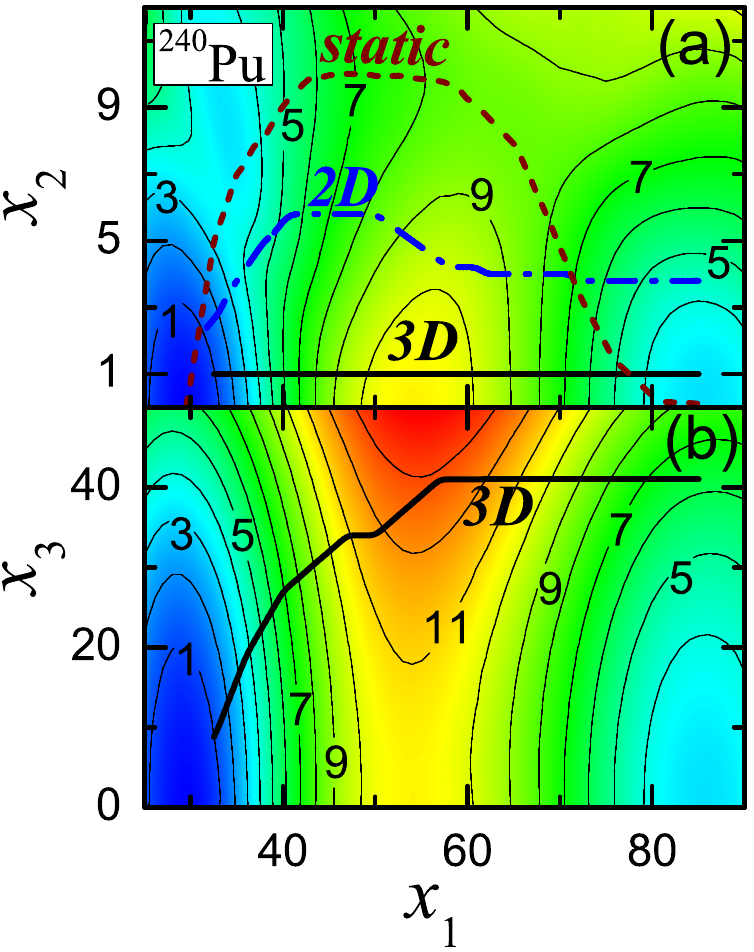}
\caption[C1]{\label{plot6}
(Color online) Similar as in Fig.~\ref{plot4} but for $^{240}$Pu. The
static SF path is marked by the
dotted line.}
\end{figure}
As seen in Fig.~\ref{plot6}, in the region of the first saddle in the static calculations,
the impact of dynamics on the
least-action pathway for $^{240}$Pu is dramatic. Compared to
static-pairing calculation, in the 2D calculations the effect of
triaxiality is significantly reduced, and in 3D calculations the axial
symmetry of the system is fully restored.

{\it Conclusions} --- In this study, we extended the self-consistent
least-action approach to the SF by considering collective coordinates
associated with pairing. Our approach takes into account  essential
ingredients impacting the SF dynamics \cite{(Neg89)}: (i) spontaneous
breaking of  mean field symmetries; (ii) diabatic configuration
changes due to level crossings; (iii) reduction of nuclear inertia
by pairing; and (iv) dynamical fluctuations governed by the
least-action principle.

We demonstrated that the SF pathways and lifetimes are significantly
influenced by the nonperturbative collective inertia and dynamical
fluctuations in shape and pairing degrees of freedom. While the
reduction of the collective action by pairing fluctuations has been
pointed out in earlier works
\cite{(Mor74),(Laz87),(Sta85),(Sta89),(Loj99)} and also very recently
in a self-consistent approach~\cite{(Giu14)}, our work shows that
pairing  dynamics can profoundly impact penetration probability,
that is, effective fission barriers, by restoring symmetries
spontaneously broken in a static approach.

Our calculations for $^{264}$Fm and $^{240}$Pu show that the
dynamical coupling between shape and pairing degrees of freedom can
lead to a dramatic departure from the standard static picture based
on saddle points obtained in static mean-field calculations. In
particular, for $^{240}$Pu, pairing fluctuations restore the axial
symmetry around the fission barrier, which in the static approach is
broken spontaneously. The examples presented in this work, in
particular in Figs.~\ref{plot5} and \ref{plot6}, illustrate how
limited is the notion of fission barrier.

The future improvements, aiming at systematic comparison with
experiment, will include: the full Adiabatic Time Dependent HFB
treatment of collective inertia, adding reflection asymmetric
collective coordinates, and employing   energy density functionals
optimized for fission \cite{(Kor12)}. The work along all these lines
is in progress.

\begin{acknowledgments}
Discussions with G.F. Bertsch, K. Mazurek, and N. Schunck are gratefully acknowledged. This study was initiated during the  Program INT-13-3 ``Quantitative
Large Amplitude Shape Dynamics: fission and heavy ion fusion" at the
National Institute for Nuclear Theory in Seattle. This material is
based upon work supported by the U.S. Department of Energy, Office of
Science, Office of Nuclear Physics under Award Numbers No.\
DE-FG02-96ER40963 (University of Tennessee) and No.\ DE-SC0008499
(NUCLEI SciDAC Collaboration); by the NNSA's Stewardship Science
Academic Alliances Program under Award No. DE-FG52-09NA29461 (the
Stewardship Science Academic Alliances program); by the Academy of
Finland and University of Jyv\"askyl\"a within the FIDIPRO programme;
and by the Polish National Science Center under Contract
Nos.~2012/07/B/ST2/03907. An award of
computer time was provided by the National Institute for
Computational Sciences (NICS) and the Innovative and Novel
Computational Impact on Theory and Experiment (INCITE) program using
resources of the OLCF facility.
\end{acknowledgments}

\bibliographystyle{apsrev4-1}
\bibliography{ref}

\end{document}